\documentclass[prl,twocolumn,superscriptaddress,showpacs]{revtex4}
\usepackage{amsmath}
\usepackage{amsfonts}
\usepackage{amssymb}
\usepackage{graphicx}

\begin{document}

\title{Quantum Entanglement in Time}
\author{{\v C}aslav Brukner}
\email{caslav.brukner@univie.ac.at} \affiliation{Optics Section,
The Blackett Laboratory, Imperial College, Prince Consort Road,
London, SW7 2BW, United Kingdom} \affiliation{Institut f\"ur
Experimentalphysik, Universit\"at Wien, Boltzmanngasse 5, A--1090
Wien, Austria}
\author{Samuel Taylor}
\email{samuel.taylor@ic.ac.uk} \affiliation{Optics Section, The
Blackett Laboratory, Imperial College, Prince Consort Road,
London, SW7 2BW, United Kingdom}
\author{Sancho Cheung}
\email{shu.cheung@ic.ac.uk} \affiliation{Optics Section, The
Blackett Laboratory, Imperial College, Prince Consort Road,
London, SW7 2BW, United Kingdom}
\author{Vlatko Vedral}
\email{v.vedral@ic.ac.uk} \affiliation{Optics Section, The
Blackett Laboratory, Imperial College, Prince Consort Road,
London, SW7 2BW, United Kingdom}
\date{\today}

\begin{abstract}

The temporal Bell inequalities are derived from the assumptions of
realism and locality in time. It is shown that quantum mechanics
violates these inequalities and thus is in conflict with the two
assumptions. This can be used for performing certain tasks that
are not possible classically. Our results open up a possibility
for introducing the notion of entanglement in time in quantum
physics.

\end{abstract}

 \maketitle

Conceptually, as well as mathematically, space and time are
differently described in quantum mechanics. While time enters as
an external parameter in the dynamical evolution of a system,
spatial coordinates are regarded as quantum-mechanical
observables. Moreover, spatially separated quantum systems are
associated with the tensor product structure of the Hilbert
state-space of the composite system. This allows a composite
quantum system to be in a state that is not separable regardless
of the spatial separation of its components. We speak about
\emph{entanglement in space}. On the other hand, time in quantum
mechanics is normally regarded as lacking such a structure.

Entanglement in space displays one of the most interesting
features of quantum mechanics, often called quantum nonlocality.
\emph{Locality in space} and {\it realism} impose constraints -
Bell's inequalities~\cite{bell} - on certain combinations of
correlations for measurements of spatially separated systems,
which are violated by quantum mechanics. Furthermore, entanglement
in space is considered as a resource that allows powerful new
communication and computational tasks that are not possible
classically~\cite{nielsen}.

Because of different roles time and space play in quantum theory
one could be tempted to assume that the notion of
\emph{``entanglement in time''} cannot be introduced in quantum
physics. In this letter we will investigate this question and we
will find that this is not the case.

We will explicitly derive \emph{temporal Bell's inequalities} (the
notion of temporal Bell's inequalities was first introduced by
Leggett and Garg~\cite{leggett} in a different context; see
discussion below) in analogy to the spatial ones. They are
constraints on certain combinations of temporal correlations for
measurements of a \emph{single} quantum system, which are
performed at \emph{different} times. We explicitly show that
quantum mechanics violates these inequalities. While
mathematically two-fold correlations in space and in time are
equivalent, the general spatial and temporal $m$-fold correlations
can have completely different features. On one hand, every
$m$-fold temporal correlation is decomposable into two-fold
correlations, so that no Bell's inequalities that detect genuine
$m$-fold $(m\!>\!2)$ nonseparability~\cite{svetlichny} can be
violated. On the other hand, and in apparent contradiction with
this, the temporal correlations may be stronger than the spatial
ones in a certain sense. Finally, we show that entanglement in
time can save on the size of classical memory required in certain
computational problems beyond the classical limits.

The temporal Bell's inequalities are derived from the following
two assumptions~\cite{freewill}: (a) {\emph{Realism:} The
measurement results are determined by "hidden" properties the
particles carry prior to and independent of observation, and (b)
\emph{Locality in time:} The results of measurement performed at
time $t_2$ are independent of any measurement performed at some
earlier or later time $t_1$. It should be noted that in contrast
to spatial correlations, where the special theory of relativity
can be invoked to ensure locality in space, no such principle
exists to ensure locality in time for temporal correlations.
Nevertheless, it is meaningful to ask whether or not the
quantum-mechanical predictions are compatible with the assumptions
(a) and (b). Ultimately we expect to learn more about the relation
between the structure of space and time and the abstract formalism
of quantum theory.


We comment on important related works~\cite{leggett,paz}. While
there temporal Bell's inequalities are for histories, our
inequalities are for predetermined measurement values. Also, there
the observer measures a \emph{single} observable having a choice
between \emph{different} times $t_i$ of measurement (the times
$t_i$ play the role of measurement settings), whereas in our case
at any given time $t_i$ the observer has a choice between
\emph{different} measurement settings. We will see that the
possibility of the observer to choose between different
observables is decisive for our new computational task that is not
possible classically.

We shall now derive the temporal analog of the
Clauser-Horne-Shimony-Holt (CHSH) inequality~\cite{chsh} from the
assumptions (a) and (b). Consider an observer and allow her to
choose at time $t_i$ to measure between two dichotomic
observables, determined by some parameters $\vec{n}^1_i$ and
$\vec{n}^2_i$. The assumptions (a) and (b) imply existence of
numbers $A_{t_i}(\vec{n}^1_i)$ and $A_{t_i}(\vec{n}^2_i)$ each
taking values either +1 or -1, which describe the predetermined
result of the measurement performed at time $t_i$ of the
observable defined by $\vec{n}^1_i$ and $\vec{n}^2_i$,
respectively.

In a specific sequence of $m$ measurements performed at the set of
times $t_1,...,t_m$, the correlations between $m$ observations are
given by the product $\prod_{j=1}^m A_{t_j}(\vec{n}^{k_j}_j)$,
with $k_j\!=\!1,2$. The temporal correlation function is then the
average over many runs of the sequence of measurements, as given
by $E(\vec{n}^{k_1}_{1},...,\vec{n}^{k_m}_{m})\!=\!\langle
\prod_{j=1}^m A_{t_j}(\vec{n}^{k_j}_i) \rangle_{avg}$.

In what follows we will consider only correlations for
measurements performed at two different times $t_1$ and $t_2$. To
avoid too many indices we introduce a new notation for
predetermined values: $A^1_{t_1}$ and $A^2_{t_1}$ stand for the
measurement results at time $t_1$ for the observables $\vec{a}_1$
and $\vec{a}_2$, and $B^1_{t_2}$ and $B^2_{t_2}$ are the results
at time $t_2$ for $\vec{b}_1$ and $\vec{b}_2$, respectively. The
following algebraic identity holds for the predetermined values: $
A^1_{t_1} (B^1_{t_2} + B^2_{t_2}) + A^2_{t_1} (B^1_{t_2} -
B^2_{t_2}) = \pm 2.$ After averaging this expression over many
runs of the sequence of measurements, one obtains the temporal
CHSH inequality
\begin{equation}
B \equiv |E(\vec{a}_1,\vec{b}_1) + E(\vec{a}_1,\vec{b}_2) +
E(\vec{a}_2,\vec{b}_1) - E(\vec{a}_2,\vec{b}_2)| \leq 2
\label{chsh}
\end{equation}
in analogy to the spatial one. We call expression $B$ on the
left-hand side of ineq. (\ref{chsh}) the Bell expression. Note
that probability distribution in phase space in classical
mechanics satisfy this inequality.

We will now calculate the temporal correlation function for
consecutive measurements of a single qubit. An arbitrary mixed
state of a qubit can be written as $ \rho = \frac{1}{2}
(\mathbb{I} + \vec{r}\cdot \vec{\sigma})$, where $\mathbb{I}$ is
the identity operator, $\vec{\sigma}\equiv
(\sigma_x,\sigma_y,\sigma_z)$ are the Pauli operators for three
orthogonal directions $x$, $y$ and $z$, and $\vec{r}\equiv
(r_x,r_y,r_z)$ is the Bloch vector with the components
$r_i\!=\!\mbox{Tr}(\rho \sigma_i)$. Here "$\cdot$" denotes the
ordinary scalar product in a three-dimensional Euclidian space.

Suppose that the measurement of the observable $\vec{\sigma} \cdot
\vec{a}$ is performed at time $t_1$, followed by the measurement
of $\vec{\sigma} \cdot \vec{b}$ at $t_2$, where $\vec{a}$ and
$\vec{b}$ are directions at which spin is measured. The quantum
correlation function is given by $ E_{QM}(\vec{a},\vec{b}) =
\sum_{k,l=\pm 1} k \cdot l \cdot \mbox{Tr}(\rho P^{k}_{\vec{a}})
\cdot \mbox{Tr}(P^{k}_{\vec{a}} P^{l}_{\vec{b}}),$ where, e.g.,
$P^{k}_{\vec{a}}$ is the projector onto the subspace corresponding
to the eigenvalue $k\!=\!\pm 1$ of the spin along $\vec{a}$. Here
we use the fact that after the first measurement the state is
projected on the new state $P^{k}_{\vec{a}}$. Therefore, the
probability to obtain the result $k$ in the first measurement and
$l$ in the second one is given by $\mbox{Tr}(\rho P^{k}_{\vec{a}})
\cdot \mbox{Tr}(P^{k}_{\vec{a}} P^{l}_{\vec{b}})$. Using
$P^{k}_{\vec{a}}= \frac{1}{2} ( \mathbb{I} + k \vec{\sigma} \cdot
\vec{a})$ and $\frac{1}{2} \mbox{Tr} [(\vec{\sigma} \cdot \vec{a})
(\vec{\sigma} \cdot \vec{b})]= \vec{a} \cdot \vec{b}$ one can
easily show that the quantum correlation function can simply be
written as
\begin{equation}
E_{QM}(\vec{a},\vec{b}) = \vec{a} \cdot \vec{b}.\label{eqm}
\end{equation}
It is remarkable that in contrast to the spatial correlation
function the temporal one (\ref{eqm}) does not dependent of the
initial state $\rho$. We note that our derivation of Eq.
(\ref{eqm}), if one adopts Heisenbergs picture, also includes the
cases where the system evolves under arbitrary unitary
transformation between the two measurements.

We shall now show that quantum mechanics violates the temporal
CHSH inequality and is thus in conflict with the assumptions (a)
and (b). We compute the quantum value $B_{QM}$ for the Bell
expression where the observer can choose between observables
$(\vec{\sigma} \cdot \vec{a}_1)$ and $(\vec{\sigma} \cdot
\vec{a}_2)$ at time $t_1$ and between $(\vec{\sigma} \cdot
\vec{b}_1)$ and $(\vec{\sigma} \cdot \vec{b}_2)$ at $t_2$. We
obtain
\begin{equation}
B_{QM} = |\vec{a}_1 \cdot (\vec{b}_1+\vec{b}_2) + \vec{a}_2 \cdot
(\vec{b}_1-\vec{b}_2)|.
\end{equation}
The maximal violation of the temporal CHSH inequality is achieved
for the choice of the measurement settings: $\vec{a}_1 \!=\!
\frac{1}{\sqrt{2}} (\vec{b}_1 + \vec{b}_2) $ and $\vec{a}_1 \!=\!
\frac{1}{\sqrt{2}} (\vec{b}_1 - \vec{b}_2)$ and is equal to
$2\sqrt{2}$. This can be called the temporal Cirel'son
bound~\cite{cirelson}.

We now consider the situation of three consecutive observations.
We are still interested in two-fold correlations for two
measurements performed, say, at times $t_1$ and $t_3$, but where
an additional measurement is performed at time $t_2$ lying between
$t_1$ and $t_3$ ($t_1\!<\!t_2\!<\!t_3$). Suppose that the three
measurements are: $(\vec{\sigma} \cdot \vec{a})$ at time $t_1$,
$(\vec{\sigma} \cdot \vec{b})$ at $t_2$ and $(\vec{\sigma} \cdot
\vec{c})$ at $t_3$. Applying the similar method as used for
computing Eq. (\ref{eqm}) one obtains:
\begin{eqnarray}
E_{QM}(\vec{a},\vec{c}) &=& \hspace{-0.3cm} \sum_{k,l,s=\pm 1}
\hspace{-0.2cm} k \cdot s \cdot \mbox{Tr}(\rho P^{k}_{\vec{a}})
\cdot \mbox{Tr}(P^{k}_{\vec{a}} P^{l}_{\vec{b}}) \cdot
\mbox{Tr}(P^{l}_{\vec{b}} P^{s}_{\vec{c}}) \nonumber \\
&=& (\vec{a} \cdot \vec{b}) (\vec{b} \cdot \vec{c}) \label{dis}
\end{eqnarray}
for the correlation for measurements performed at times $t_1$ and
$t_3$. One can convince oneself that the correlation function
(\ref{dis}) for a given measurement performed at $t_2$ cannot
violate the temporal CHSH inequality for measurements at $t_1$ and
$t_3$. Therefore, any measurement performed at time $t_2$
``disentangles'' events at times $t_1$ and $t_3$ if
$t_1\!<\!t_2\!<\!t_3$.

It should be noted that the temporal correlation as given by Eq.
(\ref{eqm}) (with a minus sign in front) can also be obtained for
results of the consecutive measurements of two qubits that are in
the maximally entangled state (singlet). We will see, however,
that the equivalence between spatial and temporal correlations is
not a general feature, but rather a peculiarity of the two-fold
correlations. In fact, the quantum correlation
$E_{QM}(\vec{n}_{1},...,\vec{n}_{m})$ for measurements performed
at $m$ instances of time $t_1,...,t_m$ is decomposable into a
product of two-fold temporal correlations of the type (\ref{eqm}).
One obtains
\begin{eqnarray}
\lefteqn{E_{QM}(\vec{n}_{1},...,\vec{n}_{m}) =} \\
&& \hspace{-0.8cm} \sum_{s_1,...,s_m=\pm 1} \hspace{-0.3cm}
s_1...s_m \mbox{Tr}(\rho P^{s_1}_{\vec{n}_1})
\mbox{Tr}(P^{s_1}_{\vec{n}_1} P^{s_2}_{\vec{n}_2}) ...
\mbox{Tr}(P^{s_{m-1}}_{\vec{n}_{m-1}} P^{s_m}_{\vec{n}_m})
\nonumber
\\
&=& E_{QM}(\vec{n}_{1},\vec{n}_{2}) \cdot
E_{QM}(\vec{n}_{3},\vec{n}_{4}) \cdot ...\cdot
E_{QM}(\vec{n}_{m-1},\vec{n}_{m}), \nonumber
\end{eqnarray}
where we assume that $m$ is even for simplicity. This implies
that, in contrast to general spatial correlation, correlation in
time is partially separable, i.e. any $m$ events in time are
composed of sets of pairs of events which may be correlated in any
way (e.g. entangled) within the pair but which are uncorrelated
with respect to the events from other pairs. Consequently, the
Bell-type inequalities that detect genuine $m$-fold
nonseparability~\cite{svetlichny} are satisfied by the temporal
correlations for all $m\!>\!2$.

One can understand this in the following way. With the only
exception being when the system is in an eigenstate of the
measured observable, the set of future probabilistic predictions
specified by the new projected state is indifferent to the
knowledge collected from the previous measurements in the whole
history of the system. One has only correlations between the state
preparation, which also can be considered as a measurement of the
system at time $t_i$, and the measurement performed at the next
time $t_{i+1}$. A related view was held by Pauli~\cite{pauli} who
wrote (see~\cite{translation} for translation): ``Bei Ubestimmheit
einer Eigenschaft eines Systems bei einer bestimmten Anordnung
(bei einem bestimmten Zustand des Systems) vernichtet jeder
Versuch, die betreffende Eigenschaft zu messen, (mindestend
teilweise) den Einflu{\ss} der fr\"{u}heren Kenntnisse von System
auf die (eventuell statistischen) Aussagen \"{u}ber sp\"{a}tere
m\"{o}gliche Messungsergebnisse.''

Although in contrast to correlations in space there are no genuine
multi-mode correlations in time, we will see that temporal
correlations can be stronger than spatial ones in a certain sense.
We denote by $\max{[B^{space}_{QM}(i,j)]}$ the maximal value of
the Bell expression for qubits $i$ and $j$. Scarani and
Gisin~\cite{scarani} found an interesting bound that holds for
arbitrary state of three qubits:
\begin{equation}
\max{[B^{space}_{QM}(1,2)]} + \max{[B^{space}_{QM}(2,3)]} \leq 4.
\label{qspace}
\end{equation}
Physically, this means that no two pairs of qubits of a
three-qubit system can violate the CHSH inequalities
simultaneously. This is because if two systems are highly
entangled, they cannot be entangled highly to another systems. Let
us denote by $\max{[B^{time}_{QM}(1,2)]}$ the maximal value of the
Bell expression for two consecutive observations of a single qubit
at times $i$ and $j$. Since temporal two-fold correlations
(\ref{eqm}) do not depend on the initial state one can simply
combine them to obtain
\begin{equation}
\max{[B^{time}_{QM}(1,2)]} + \max{[B^{time}_{QM}(2,3)]} \leq 4
\sqrt{2}. \label{qtime}
\end{equation}
Thus, although there are no genuine 3-fold temporal correlations,
a specific combination of two-fold correlations can have values
that are not achievable with correlations in space for any 3-qubit
system. In fact, one would need \emph{two} pairs of maximally
entangled two-qubit states (two e-bits) to achieve the bound in
(\ref{qtime}). Also note that the local realistic bound is 4,
which is equal to the bound in (\ref{qspace}) but lower than the
one in (\ref{qtime}). Similar conclusion can be obtained for the
sum of $m$ Bell's expressions.

We now show that entanglement in time can be used to perform
certain tasks that are not possible classically. With
``classically'' we mean here compatible with the assumptions (a)
and (b). Our tasks are temporal analogue of communication
complexity problems~\cite{yao,buhrman,brukner,brassard} with space
and time interchanging roles. Consider a party who receives a set
of input data $z_1,...,z_m$ at different times $t_1,...,t_m$,
respectively. The goal for her is to determine the value of a
certain function $f(z_1,...,z_m)$ that depends on all data. During
the protocol she is allowed to use random strings that are
\emph{classically} correlated in time, which might improve the
success of the protocol. Obviously, if the party has enough memory
to store all $m$ inputs, she can compute the function with
certainty after receiving all inputs. We will consider the
problem: What is the highest possible probability for the party to
arrive at the correct value of the function if only a
\emph{restricted} amount of the (classical) memory is available?
We will show that there are functions $f$ for which the party can
increase the success rate, if she uses entanglement in time rather
than random strings correlated classically in time.

\begin{figure}
\centering
\includegraphics[angle=0,width=8.6cm]{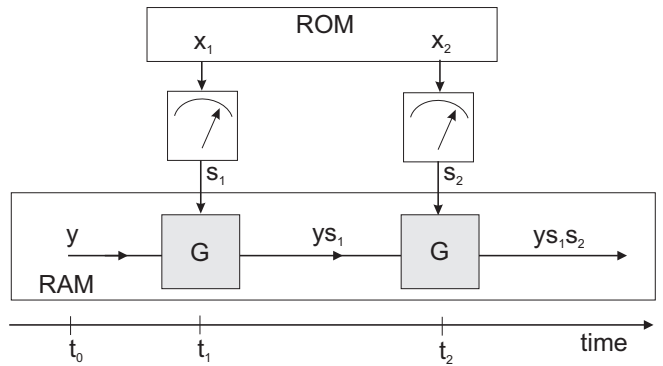}
\caption{Quantum protocol for computing the function given in Eq.
(\ref{prva}). The size of the RAM is restricted to 1 bit. A party
receives input bits $y,x_1,x_2$ at times $t_0,t_1,t_2$,
respectively (e.g. from the computer ROM). She has a qubit system
and a suitable device for measuring two-valued observables. The
input $y$ is feed into RAM. Depending on the specific value $x_i$,
$i=1,2$ of her input, the party chooses to measure between two
different observables. Applying the gate $G$ the result $s_i$ of
the measurement is multiplied by the previous bit value of the
RAM. The final output is $y \cdot s_1 \cdot s_2$.} \label{memory}
\hspace{-0.4cm}
\end{figure}

We will consider a function introduced previously in the framework
of communication complexity~\cite{buhrman,brukner}. Imagine that
the party receives bit input $y \!\in\! (-1,1)$ at time $t_0$,
$x_1 \!\in\! (0,1)$ at $t_1$, and finally $x_2 \!\in\! (0,1)$ at
$t_2$. One can imagine that she has an access to her ROM
(read-only memory - computer memory whose contents can be only be
read out) where the inputs are stored only at times $t_i$,
$i=0,1,2$. Her goal is to compute the function
\begin{equation}
f(y,x_1,x_2) = y \cdot (-1)^{x_1 x_2} \label{prva}
\end{equation}
with as high probability as possible, while having only amount of
\emph{1 bit of memory (RAM)} available. In other words, the size
of her RAM (random-access memory - computer memory which can be
used to perform necessary tasks) is restricted to 1 bit.

We will first present a quantum protocol and then show that it is
more efficient than any classical one. The party receives the
input $y$ at time $t_0$ and feeds it into her RAM as shown in
Fig.~\ref{memory}. If at time $t_1$ she receives $x_1\!=\!0$, she
will measure an observable $(\vec{\sigma} \cdot \vec{a}_1)$ on her
qubit. For $x_1\!=\!1$, she will measure $(\vec{\sigma} \cdot
\vec{a}_2)$. The actual value $\pm 1$ obtained in the measurement
is denoted by $s_1$. The party uses a multiplication gate $G$ to
multiply the value $y$ of the bit in the RAM with the measurement
result $s_1$. She obtains $y \!\cdot \!s_1$ as the new value in
her RAM. The same strategy is repeated also at time $t_2$ with the
measurement of observable $(\vec{\sigma} \cdot \vec{b}_1)$ or
$(\vec{\sigma}\cdot \vec{b}_2)$ and the actual result denoted by
$s_2$. As the final output the party obtains $y \cdot s_1 \cdot
s_2$.

The probability of the success of the quantum protocol is equal to
the probability $P$ for the product $s_1 \cdot s_2$ to be equal to
$(-1)^{x_1 x_2}$. This probability can be written as
\begin{eqnarray}
P &=& \frac{1}{4} [P_{\vec{a}_1,\vec{b}_1}(s_1s_2 \!=\!1) +
P_{\vec{a}_1,\vec{b}_2}(s_1s_2\! =\!1) \nonumber \\ & +&
P_{\vec{a}_2\vec{b}_1} (s_1s_2 \!=\!1) + P_{\vec{a}_2,\vec{b}_2}
(s_1s_2\! = \!-1)], \label{tigar}
\end{eqnarray}
where, e.g., $P_{\vec{a}_1,\vec{b}_2}(s_1s_2\!=\!1)$ is the
probability that $s_1s_2 \!=\!1$ if the party receives input value
$x_1\!=\!0$ at $t_1$ and $x_2\!=\!1$ at $t_2$. All four possible
input combinations occur with the same probability $1/4$.

It is important to see that the probability of success can be
expressed as $P\!=\!\frac{1}{4} B$, where $B$ is the Bell
expression given in (\ref{chsh}). On the other hand, a classical
protocol, that is any protocol that is compatible with the
assumptions (a) and (b), can be understood as exploiting a
realistic and local-in-time model of the quantum protocol. This
implies that the probability of the success in any classical
protocol~\cite{comment} is bounded, i.e. $P_C \leq 3/4 \!=
\!75\%$. The quantum protocol will have higher success if and only
if the choice of the pair of measurements at time $t_1$ and $t_2$
violates the temporal CHSH inequality. With the optimal choice of
the measurements the probability of success is $P_Q\!=\!85\%$.
Note that to achieve this success rate classically one has to use
at least the size of \emph{two} bits of the memory (RAM).

We note that one can also construct tasks whose quantum solutions
would exploit violation of the classical bound for the expression
(\ref{qtime}). Not only that those tasks would not be possible
classically but they also could not be efficiently performed with
spatial entanglement without additional resources (e-bits).
Finally, we note that quantum communication complexity protocols
that are not based on entanglement but on exchange of
qubits~\cite{yao,brassard} can also be reformulated within the
framework of our tasks. Note, however, that here we are interested
in exchanging classical bits rather than qubits. This exploits
entanglement in time as a resource for increasing capacities of
classical computational devices.

One related issue we have not explored in this paper is that of
mathematically describing time by associating a tensor product
structure to a sequence of time instances. This seems to be
necessitated by our notion of entanglement in time. Finkelstein
was first to consider something similar when he introduced
quantised instances of time called chronons~\cite{finkelstein}.
More recently, Isham, within the framework of consistent
histories, explored the same possibility~\cite{isham}. It is clear
from our work, however, that it is very difficult to extend the
tensor product structure beyond the two neighbouring instances in
time without altering the basic principles of quantum mechanics.
In fact, one of the features of entanglement in time is exactly a
consequence of this difficulty: two maximally entangled events can
still be maximally entangled to two other events in time (a
principle we may call ``polygamy'' of entanglement in time). This
is in contrast to the spatial entanglement which can only be
``monogamous''~\cite{bennett}. The difference between the spatial
and temporal structure may ultimately be fundamental, or it may be
an indication that we need a deeper theory in which the two need
to be treated on a more equal footing (quantum field theory does
not suffice in this sense). Either way, it appears that the next
step should lie in exploring the consequences of combining
entanglement in space and time in order to study how they relate
to each other.

\begin{acknowledgments}

{\v C}.B. has been supported by the Marie Curie Fellowship,
Project No. 500764. V.V. has been supported by Engineering and
Physical Sciences Research Council, the European Commission and
Elsag-spa company.

\end{acknowledgments}


\begin{thebibliography}{99}
\bibitem{freewill} The assumption of ``free-will'' - the freedom of
the observer to choose the measurement setting - is implicitely
assumed.
\bibitem{translation} ``In the case of
undefiniteness of a property of a system for a certain arrangement
(with certain state of the system) any attempt to measure that
specific property destroys (at least partially) the influence of
earlier knowledge of the system on (possibly statistical)
statements about later possible measurement results.''
\bibitem{comment} One can show that even if one considers the most general classical protocols
$P_C$ cannot exceed 75\%. The proof is along the similar lines as
the one in Ref.~\cite{brukner}.
\bibitem{bennett} As far as we are aware Bennett was the first to introduce the
notion of ``monogamy'' of entanglement.
\end{thebibliography}
\end{document}